\documentclass[12pt,preprint]{aastex}

\newcommand\unit[1]{\ \mathrm{#1}}
\def\My{\unit{My}}
\def\Mjup{\unit{M_{jup}}}
\def\AU{\unit{AU}}
\def\ip{{\mathrm{ip}}}
\def\op{{\mathrm{op}}}
\def\AU{\unit{AU}}

\def\My{\unit{My}}

\def\Mearth{\unit{M}_\earth}

\begin{document} 
\title{Prospects for the habitability of OGLE-2006-BLG-109L}
\author{Renu Malhotra\altaffilmark{1} and David A. Minton\altaffilmark{1}}
\altaffiltext{1}{Department of Planetary Sciences, University of Arizona, Tucson, AZ 85721; renu@lpl.arizona.edu, daminton@lpl.arizona.edu}

\begin{abstract}
The extrasolar system OGLE-2006-BLG-109L is the first multiple-planet system to be discovered by gravitational microlensing~\citep{Gaudi:2008Science}; the two large planets that have been detected have mass ratios, semimajor axis ratios, and equilibrium temperatures that are similar to those of Jupiter and Saturn; the mass of the host star is only $0.5 M_\odot$, and the system is more compact than our own Solar system.  We find that in the habitable zone of the host star, the two detected planets resonantly excite large orbital eccentricities on a putative earth-mass planet, driving such a planet out of the habitable zone. We show that an additional inner planet of $\gtrsim0.3\Mearth$ at $\lesssim0.1$ AU would suppress the eccentricity perturbation and greatly improve the prospects for habitability of the system.  Thus, the planetary architecture of a potentially habitable OGLE-2006-BLG-109L planetary system -- with two ``terrestrial'' planets and two jovian planets -- could bear very close resemblance to our own Solar system.
\end{abstract} 



\section{Introduction}\label{s:Introduction}
It is possible that the OGLE-2006-BLG-109L system harbors other planets, including earth-like planets, that are beyond the sensitivity of the microlensing observations.  Here we use dynamics to constrain the prospects for an earth-like habitable planet in this system.  We adopt \cite{Kasting:1993Icarus}'s definition of the habitable zone (HZ) as the region in which an Earth-like planet with an N$_2$--CO$_2$--H$_2$O atmosphere would be able to maintain liquid oceans on its surface.  For an Earth-like planet, the inner boundary of the HZ is the distance from the star below which the planet would experience a runaway greenhouse and lose its oceans, and the outer boundary is the distance from the star beyond which the planet would condense CO$_2$ clouds.  For OGLE-2006-BLG-109L, the HZ is $\sim0.25$--$0.36\AU$ from the star.  For comparison, and using the same definition, the HZ in our solar system is in the 0.95--1.37 AU heliocentric distance range.  Both these estimates are reported in \cite{Kasting:1993Icarus}, in their table III.

\section{Orbital dynamics of the known planets}\label{s:2plsectheory}
First, we consider the constraints on the orbital eccentricities of the two detected massive planets, because these will drive considerations for the dynamics of any habitable planets.   For the planetary parameters obtained from the observations, it is straightforward to obtain upper limits on the orbital eccentricities: if one orbit is nearly circular, then $e_1\lesssim 0.6$ or $e_2\lesssim 0.4$ will avoid crossing orbits (where subscripts 1 and 2 refer to the inner planet, planet b, and the outer planet, planet c, respectively); for comparable orbital eccentricities, the upper limit is $e_{1,2}\lesssim0.3$. The more stringent limits obtain from the resonance-overlap criterion to avoid strong dynamical chaos~\citep{Duncan:1989Icarus};  this also requires $e_{1,2}\lesssim0.3$.

\begin{figure}
\plotone{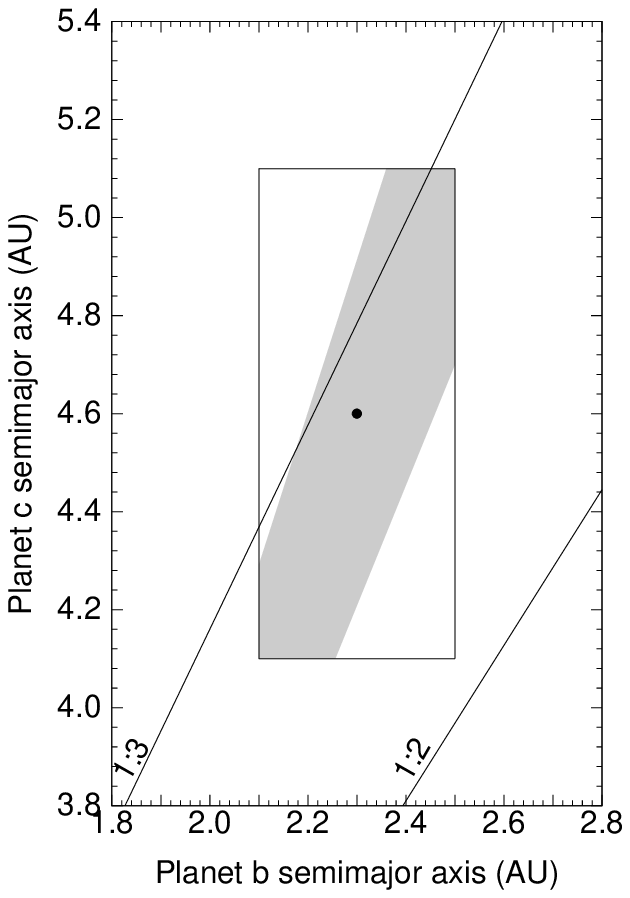}
\caption{Semimajor axes of the two observed planets in the OGLE-2006-BLG-109L system.  The rectangular box represents the uncertainty in the observations.  The shaded area is the combination of planet semimajor axes that produces a secular resonance inside the habitable zone.  Nominal locations of significant mean motion resonances are also shown.\label{f:hzresonance}}
\end{figure}

We analyzed numerically the mutual gravitational perturbations of the two planets, assuming coplanar and nearly circular orbits as initial conditions, in a grid over the range of semimajor axes spanning the observational uncertainty shown in Fig.~\ref{f:hzresonance}.  For each set of initial conditions, we used the MERCURY6 N-body integrator~\citep{Chambers:1999MNRAS} to calculate the orbital evolution of the two planets over a 1 million year duration (which is more than an order of magnitude longer than the secular timescales -- see below).  The results show that the semimajor axes remain very stable and the eccentricities exhibit only small amplitude oscillations; even in the vicinity of the 3:1 and 2:1 mean motion resonances, the maximum amplitude of the eccentricity perturbation was found to be $\sim 0.05$.   
 
The time evolution of small planetary eccentricities is described well by the classical Laplace-Lagrange linear secular perturbation theory for the eccentricity vectors, $\{h_i,k_i\} = e_i\{\sin\varpi_i,\cos\varpi_i\}$, where $\varpi_i$ is the longitude of periapse of the $i^{th}$ planet; the perturbation is governed by a set of coupled linear differential equations~\citep{Murray:1999SSD}:
\begin{equation}
	\frac{d}{dt}\left\{h_j,k_j\right\} = \sum_{i=1}^N A_{ji}\left\{k_i,-h_i\right\},\label{e:eom}
\end{equation}
(where $N=2$ for a two planet system, but the theory is generalized to any $N$).
The elements of the coefficient matrix $\mathbf{A}$ are proportional to the planetary masses and also are functions of the planetary semimajor axes.  The general solution to Eq.~(\ref{e:eom}) is:
\begin{equation}
	\left\{h_j,k_j\right\}=\sum_{i=1}^NE_j^{(i)}\left\{\sin(g_it+\beta_i),\cos(g_it+\beta_i)\right\},
	\label{e:vecsol}
\end{equation}
where the $g_i$ are the eigenvalues and $E^{(i)}$ the corresponding eigenvectors of $\mathbf{A}$.  A significant correction is owed to the averaged effects of the nearby first order 2:1 mean motion resonance (see Fig.~\ref{f:hzresonance}) which changes the coefficients by $\sim10\%$~\citep{Malhotra:1989AA}; the nearby 3:1 mean motion resonance is of second order and its effect on the secular coefficients is negligible.  In the following analysis, we include this correction which is important in locating the secular resonance accurately.  For the best-fit parameters, the two secular modes for the planets have frequencies $g_1=12.6''$/yr and $g_2=60.9''$/yr.

\section{Secular perturbations in the habitable zone}\label{s:HZsectheory}
\begin{figure}
\plottwo{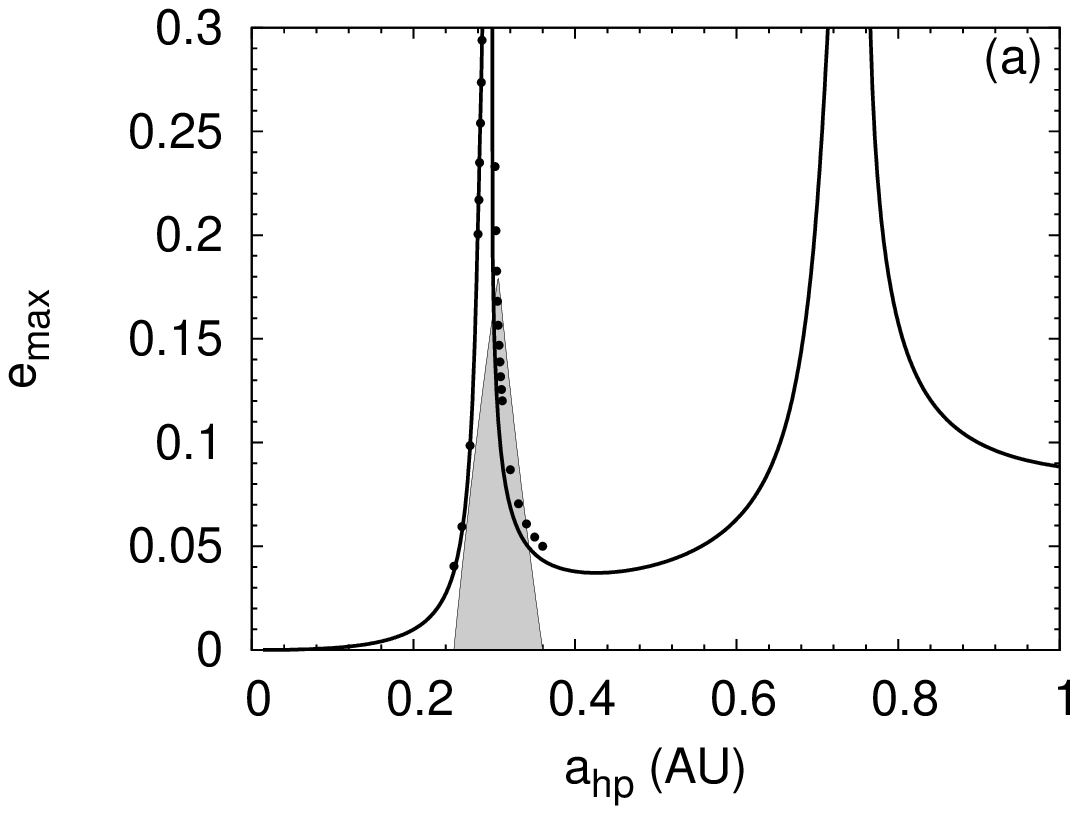}{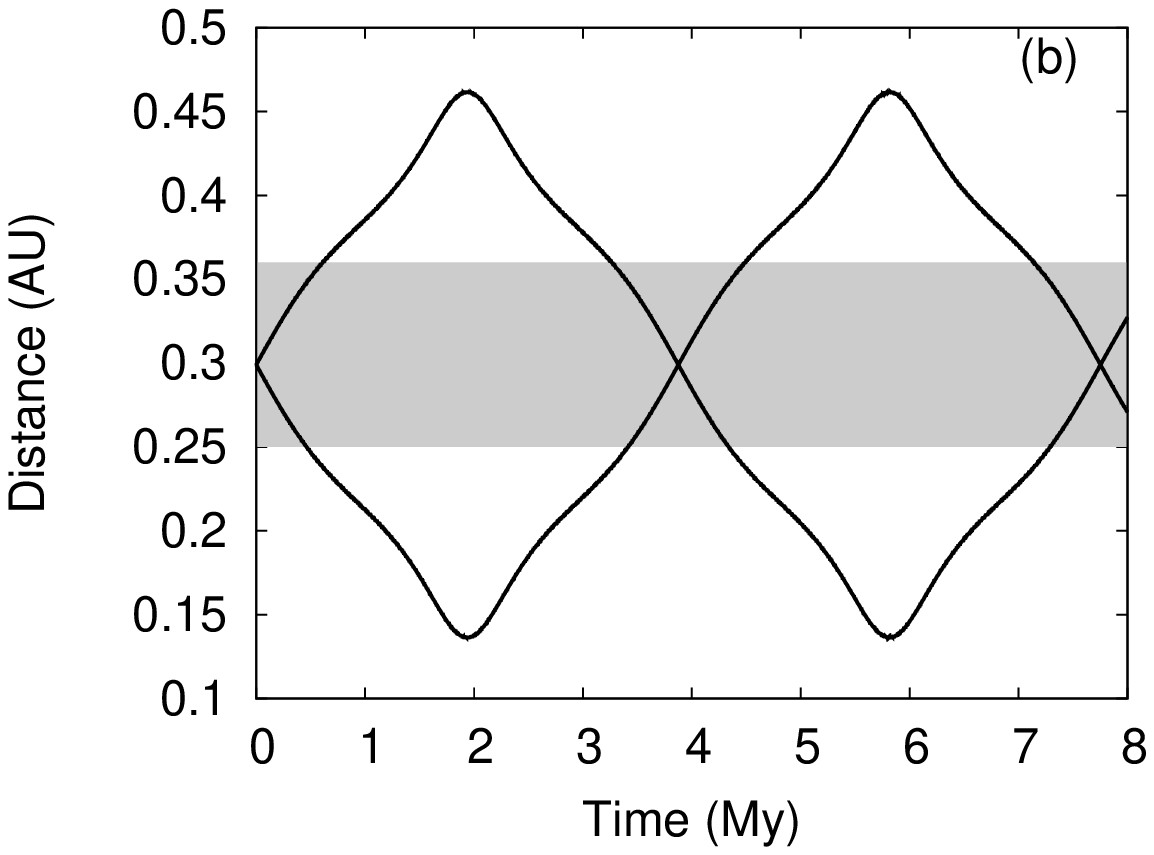}
\caption{ Eccentricity perturbations of a particle due to the secular effects of the two massive planets of OGLE-2006-BLG-109L.  The shaded region is the classical habitable zone.
{\bf (a)} Forced eccentricity of a massless test particle (solid line), and of an earth-mass planet (points).
{\bf (b)} Time evolution of the pericenter and apocenter distances of an earth-mass planet with a semimajor axis a=0.298 AU.
We used the observational best-fit semimajor axis values of $2.3\AU$ and $4.6\AU$ and mass values of $0.71\Mjup$ and $0.27\Mjup$ for planet b and c, respectively, and we assumed eccentricities of $e_1=0.03$ and $e_2=0.05$.  \label{f:innersec}} 
\end{figure}

Next, we analyze the gravitational perturbations of the two known planets on the orbital dynamics of a putative planet in the HZ.  We illustrate the effects by first considering a massless test particle: the secular perturbations forced on the test particle's eccentricity vector by the massive planets are given by~\citep{Murray:1999SSD}
\begin{equation}
	\left\{h_{\mathrm f}(t) ,k_{\mathrm f}(t)\right\}=-\sum_{i=1}^N\frac{\nu_i}{g_0-g_i}
	\left\{\sin(g_it+\beta_i),\cos(g_it+\beta_i)\right\},\label{e:forced}
\end{equation}
where $\nu_i$ are coefficients that are proportional to the planet masses and planet eccentricities, and also depend upon the semimajor axes of the planets and of the test particle; $g_0$ is the so-called free precession rate of the test particle's periastron due to the average gravitational quadrupole moment of the system.  The amplitude, $e_{\mathrm f}=\sqrt{h_{\rm f}^2+k_{\rm f}^2}$, is called the ``forced'' eccentricity because it is due to the forcing by the massive planets.  When $g_0\approx g_i$ the forced oscillations can become very large, and the test particle is said to be in a secular resonance.  
The forced eccentricity as a function of test particle semimajor axis for the inner $1\AU$ of the OGLE-2006-BLG-109L system is shown in Fig.~\ref{f:innersec}a.  
There are two strong secular resonances in this inner zone, which we label as the $\nu_1$ and the $\nu_2$ resonances, associated with the $g_1$ and $g_2$ secular modes, respectively. One of these, the $\nu_1$, is located within the habitable zone of this system, and its width is comparable to the width of the habitable zone.  Uncertainties in the planets' semimajor axes cause an uncertainty in the exact location of the secular resonance, but we find that for most of the range of planetary semimajor axes spanning the observational uncertainties, the secular resonance location remains within the HZ~(Fig.~\ref{f:hzresonance}).  

Note that in Fig.~\ref{f:innersec} (and also in Fig.~\ref{f:innersec-suppressed} below), we adopted small eccentricities, of a few percent, for the known planets; the resonant eccentricity excitation of the test particle is proportional to the amplitude of the $g_1$ secular mode of the planets, so larger planet eccentricities would lead to proportionally larger test particle excitation. For large test particle eccentricities, we must consider nonlinear corrections to the secular perturbation theory which will limit the resonant eccentricity excitation~\citep{Malhotra:1998ASPC}.  We calculate that the maximum eccentricity excitation with the $\nu_1$ secular resonance is $\sim0.55$ for $e_{1,2}\approx0.05$, and proportionally higher for higher values of $e_{1,2}$.  We also note that the linear secular perturbation theory is a good approximation as long as $e_{1,2}\lesssim0.3$; because the latter is also the stability limit discussed in section~\ref{s:2plsectheory}, our analysis here is valid for the allowed range of $e_{1,2}$.  Our analysis is also valid for non-coplanar planetary orbits if the non-coplanarity is not too large, inclination $\lesssim0.3$, so that the secular perturbations of eccentricities and inclinations remain de-coupled~\citep{Murray:1999SSD}.

The above calculation was done for a massless test particle, but its conclusions extend to terrestrial mass planets too.  We used numerical analysis to calculate the eccentricity excitation of an Earth-mass planet with semimajor axis in the habitable zone; the result, shown by the solid points in Fig.~\ref{f:innersec}a, follows closely the analytical result for the test particle.  For such a planet, large amplitude eccentricity oscillations occur with a period $\sim0.8$--$4\My$, and for most values of the semimajor axis in the HZ, the forced eccentricity will carry the planet out of the HZ for some portion of the planet's year during its high eccentricity phase~(Fig.~\ref{f:innersec}b). 

\section{Eliminating the secular resonance from the habitable zone}
\label{s:additionalplanets}
On the face of it, the analysis above would suggest that the prospects for habitability of OGLE-2006-BLG-109L are dim.  However, we can consider ways in which the secular resonance might be eliminated from the habitable zone. Towards this end, we note that the secular resonance location is sensitive to other masses that might be present in the system because these would affect the secular frequencies. In particular, additional low mass inner planets would change $g_0$ significantly while having little effect on $g_1$ and $g_2$.  Below, we calculate the mass, $m_\ip$, of an additional inner planet at semimajor axis $a_\ip$ inward of the HZ that would eliminate the secular resonance from the HZ.  We also consider solutions with an additional outer planet.

\noindent{\it Additional inner planet}\qquad With just the two observed planets, the free precession frequency $g_0(a)$ is a monotonically increasing function of $a$. The addition of an inner planet $m_\ip$ has the effect of changing the functional dependence of $g_0(a)$ so that it has a minimum somewhere between the location of the inner planet at $a_\ip$ and the location, $a_1$, of planet b.  We analytically obtain an expression for this minimum (which is a function of $m_\ip$ and $a_\ip$), then use the condition that $\min\{g_0\} > g_1$ to obtain the relationship between $m_\ip$ and $a_\ip$.   

\begin{figure}
\plotone{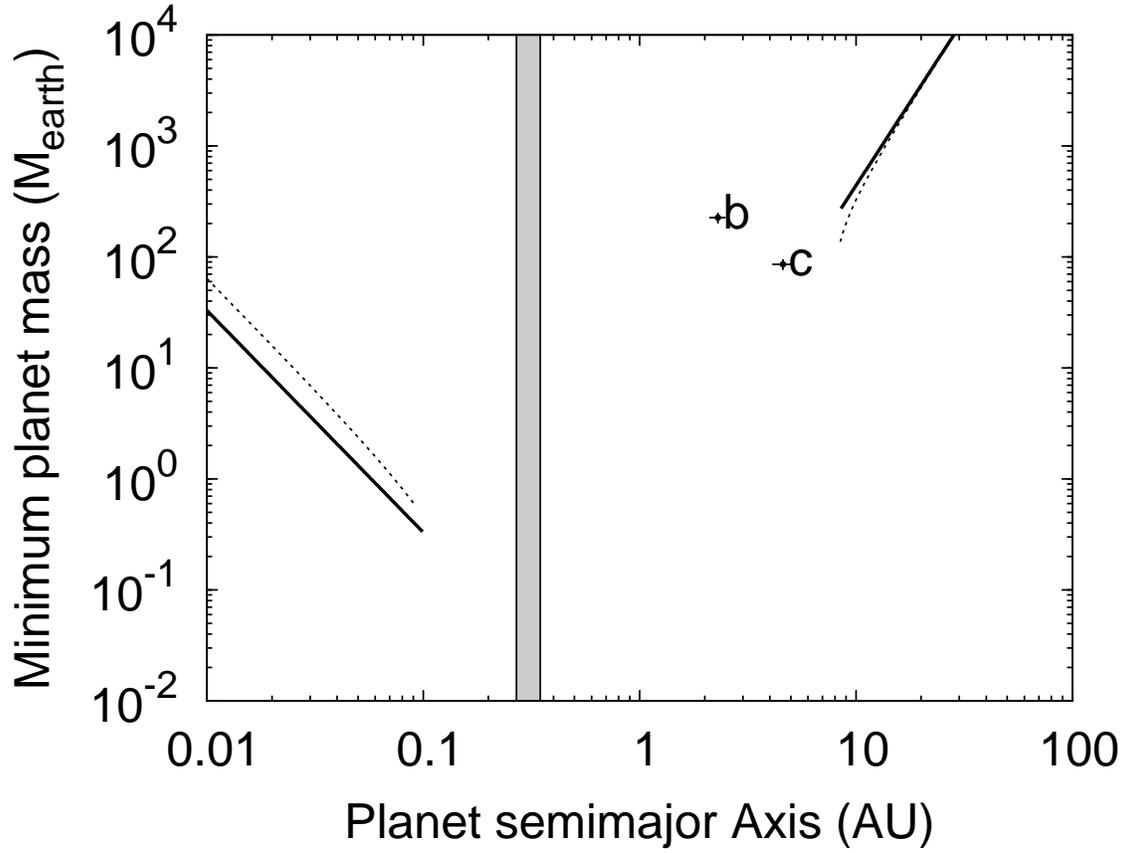}
\caption{ The solid lines are the analytical results given in Eqs.~(\ref{e:minmass-inner}) and (\ref{e:minmass-outer}).  The dashed lines are the result of solving by N-body numerical analysis for the minimum planet mass required to ensure that the forced eccentricity of a test particle is no greater than $0.1$ anywhere inside the HZ.  The shaded region is the HZ.  Shown also are the semimajor axes and masses of the two known planets, b and c.\label{f:minmass}}
\end{figure}

With three perturbing planets, the free precession frequency of a HZ planet (approximated as a test particle) is given by~\citep{Malhotra:1998ASPC}
\begin{equation}
g_0(a) = {1\over4}(Gm_*)^{1/2}\left[
{m_1\alpha_1b_{3/2}^{(1)}(\alpha_1)\over m_*a_1\sqrt{a}}
+ {m_2\alpha_2 b_{3/2}^{(1)}(\alpha_2)\over m_*a_2\sqrt{a}}
+ {m_\ip\alpha_\ip b_{3/2}^{(1)}(\alpha_\ip)\over m_*a^{3/2}} \right],
\label{e:g0}
\end{equation}
where $G$ is the universal constant of gravitation, $\alpha_\ip=a_\ip/a, \alpha_1=a/a_1, \alpha_2=a/a_2$, and $b_{3/2}^{(j)}(\alpha)$ is a Laplace coefficient~\citep{Murray:1999SSD}.  We substitute the leading order approximation, 
$b_{3/2}^{(1)}(\alpha)=3\alpha+{\cal O}(\alpha^3)$, in Eq.~\ref{e:g0}, and find 
\begin{equation}
\min\{g_0\}\simeq {15\over14}\bigg({Gm_*\over a_1^3}\bigg)^{1/2}
{m_1+(a_1/a_2)^3m_2\over m_*}\left[{7\over3}{m_\ip\over m_1+(a_1/a_2)^3m_2}\bigg({a_\ip\over a_1}\bigg)^2\right]^{3\over10}.
\label{e:g0star}
\end{equation}
Then, using the condition $\min\{g_0\} > g_1$ to eliminate the secular resonance from the HZ, we obtain
\begin{equation}
m_\ip\gtrsim0.0033\Mearth \left({a_\ip\over 1\AU}\right)^{-2}, \qquad \hbox{for $a_\ip \ll a_{\mathrm{res}}$},
\label{e:minmass-inner}
\end{equation}
where $a_{\mathrm res}\approx 0.3$~AU is the original location of the secular resonance (unperturbed by the inner planet), and we used the best-fit values of $m_*,m_1,m_2, a_1,a_2$ from \cite{Gaudi:2008Science} to obtain the numerical coefficient. The minimum inner planet mass as a function of its semimajor axis is shown in~Fig.~\ref{f:minmass}. We see, for example, that an inner planet of $\sim0.33\Mearth$ at $\sim0.1\AU$ can eliminate the secular resonance in the HZ.  A $\sim10\Mearth$ planet at $\sim0.02\AU$ (which might be called a ``hot Neptune'') would also be similarly effective.  

For a more complete analysis, we relax the test particle approximation for the HZ planet and consider the dynamics of the four planet system fully. Such a system has four secular frequencies $g_i, i=1$--$4$, and additional secular resonances become possible.  For example, with a $1 \Mearth$ HZ planet and an inner planet with $m_\ip=0.33\Mearth$ and $a_\ip=0.1\AU$, the original secular resonance is displaced well interior to the HZ, but a new secular resonance arises near the inner edge of the HZ. 
We note that our perturbation analysis is not valid when the inner planet is too close to the HZ, i.e., $a_\ip$ is in the range 0.15--0.3 AU; the perturbations are not small in this case, and numerical analysis indicates strong orbital instabilities and poor prospects for an HZ planet in this parameter regime. 

\noindent{\it Additional outer planet}\qquad 
An additional outer planet is also capable of suppressing the habitable zone secular resonance, in a slightly different way than that of the inner planet.  An outer planet, mass $m_\op$ at orbital semimajor axis $a_\op > a_2$, has the effect of increasing both the frequencies $g_0$ and $g_1$, but differentially ($g_1$ increases more than $g_0$), so that the secular resonance is displaced outward.  The increase in $g_0$ is given by 
\begin{equation}
\delta g_0\simeq{3\over4}(Gm_*)^{1/2}{m_\op a^{3/2}\over m_*a_\op^3};
\label{e:g0op}
\end{equation}
the increase in $g_1$ is given by a similar expression but with $a_1$ in place of $a$.

To displace the secular resonance out of the HZ, we require $g_1> g_0$ for all values of $a$ in the HZ.  This condition yields
\begin{equation}
m_\op\gtrsim {4\over3}{(g_{00}-g_{10})a_\op^3\over (Gm_*)^{1/2}(a_1^{3/2}-a^{3/2})} 
\simeq 0.44\Mearth \left({a_\op\over 1\AU}\right)^3,
\label{e:minmass-outer}
\end{equation}
where $g_{00}$ and $g_{10}$ are the values unperturbed by the additional planet; to evaluate the numerical coefficient on the right hand side, we used the best-fit values for the OGLE-1006-BLG-109L system, and we evaluated $g_{00}$ at the outer edge of the HZ ($a=0.36\AU$).  The forced eccentricity of a $1 \Mearth$ planet in such a system, with $m_\op=440\Mearth$ at $a_\op=10\AU$, is shown in Fig.~\ref{f:innersec-suppressed}b.
We note that an outer planet with a semimajor axis inward of $8.5\AU$ will excite a new secular resonance inside the HZ, therefore Eq.~\ref{e:minmass-outer} is valid for $a_\op\gtrsim 8.5\AU$.

The minimum mass estimates for the inner and outer planets, Eqs.~(\ref{e:minmass-inner}) and (\ref{e:minmass-outer}), are shown in Fig.~\ref{f:minmass}.  Because the outer planet's minimum mass is considerably greater than a Jupiter mass, and is also much greater than that of the detected planets, we consider it an unlikely solution.  

\section{Implications}
Dynamical considerations require the OGLE-2006-BLG-109L system to harbor at least two additional planets in order to support a potentially habitable earth-like planet.  The solution with a sub-earth mass inner planet, an $\sim$~earth-mass HZ planet and the two detected planets has an orbital structure closely resembling that of our own Solar system.  Such a system can also support planetesimal debris belts in dynamically stable zones at 0.8--1.8 AU and $\gtrsim8.5\AU$ shepherded by the major planets; these would be analogous to the Solar system's asteroid belt and Kuiper belt, respectively.  Although the solution is not unique, it is reasonable from theoretical considerations for planetary system formation~\citep[e.g.][]{Lissauer:1993ARAA}, and it is interesting to note that it would represent a closer resemblance to our own Solar system's orbital architecture than any exo-planetary system discovered thus far.  The $\sim1.5$ kpc distance to the system~\citep{Gaudi:2008Science} is too large for current radial velocity and astrometric techniques to be useful to probe this system in detail, but high-sensitivity measurements in the infrared may potentially be able to detect the interplanetary dust complex generated by any debris belts in the system.  

\acknowledgments
This research was supported in part by grants from NASA's Origins of Solar Systems Research Program
and the NASA Astrobiology Institute's node at the University of Arizona.

\bibliographystyle{apj}

\clearpage

\begin{figure} 
\plottwo{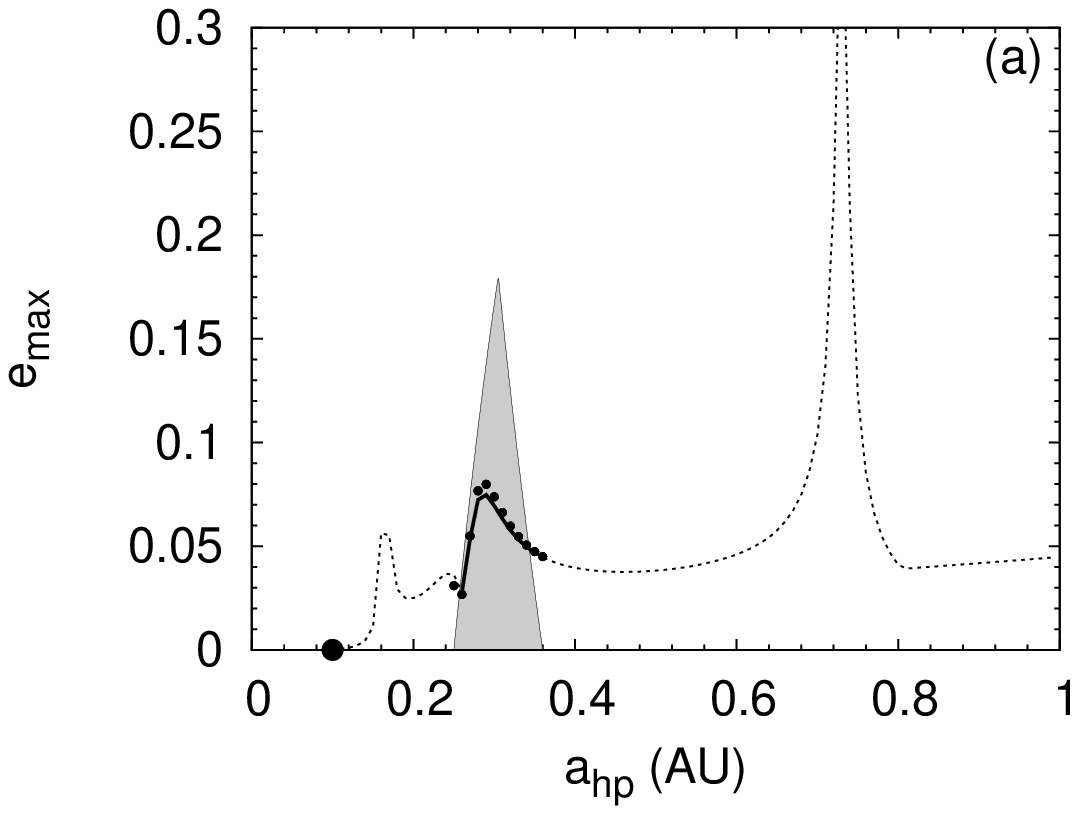}{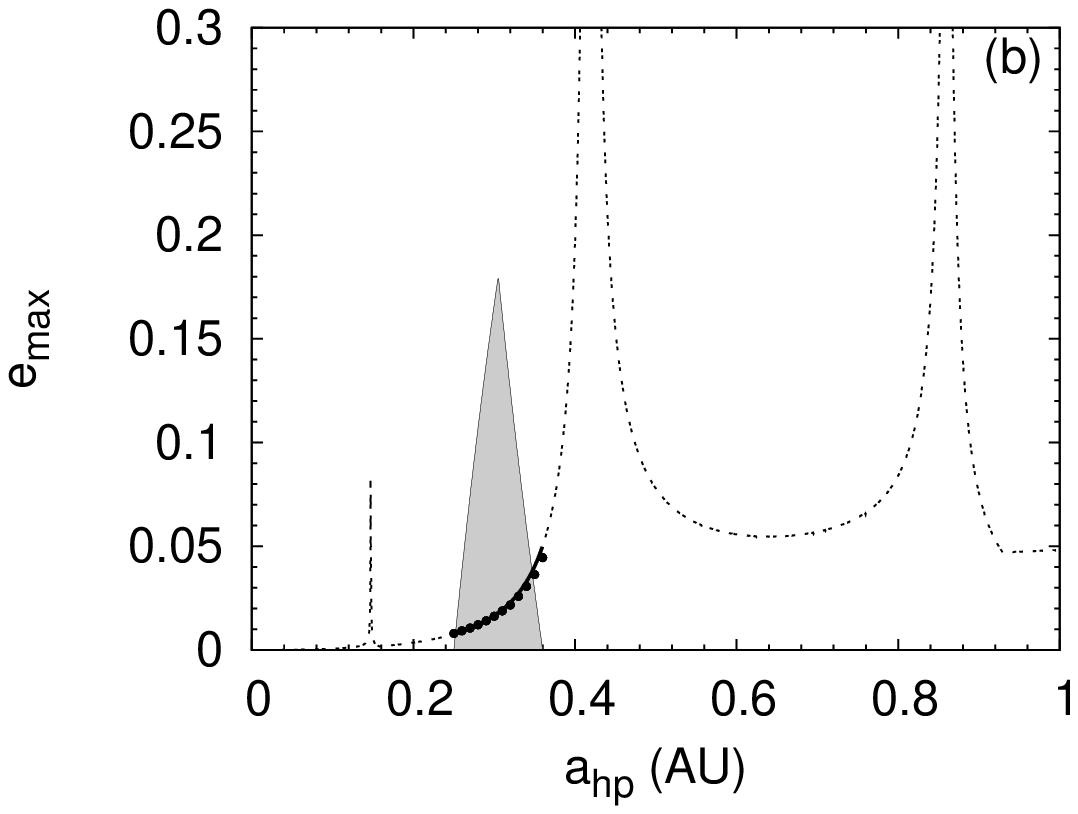}
\caption{Forced eccentricity of a $1\Mearth$ planet in the OGLE-2006-BLG-109L system with the addition of {\bf (a)} a $0.33\Mearth$ planet at $0.1\AU$ (shown as a large point),  {\bf (b)} a $440\Mearth$ planet at $10\AU$.   Dashed lines are from secular perturbation analysis; small points are from N-body numerical analysis.  Shaded area is the habitable zone.   \label{f:innersec-suppressed}}  
\end{figure} 

\end{document}